\documentclass[12pt]{JHEP3}
\usepackage{graphicx}
\usepackage{epsfig}
\usepackage{amssymb}
\usepackage{bm}
\usepackage{amsmath}
\newcommand{\bea}{\begin{eqnarray}}
\newcommand{\eea}{\end{eqnarray}}

\makeatletter
\@addtoreset{equation}{section}
\makeatother

\title{
Electromagnetic instability in holographic QCD
}
\author{
Koji Hashimoto$^{1,2,*}$, 
Takashi Oka$^{3,\#}$ and Akihiko Sonoda$^{1,\dagger}$
\\

$^1$
{\it Department of Physics, Osaka University,
Toyonaka, Osaka 560-0043, Japan}\\
$^2$
{\it Mathematical Physics Lab., RIKEN Nishina Center,
Saitama 351-0198, Japan}\\
$^3$
{\it Department of Applied Physics, University of Tokyo, 
Tokyo 113-8656, Japan}\\
E-mail: $^*$ \email{koji(at)phys.sci.osaka-u.ac.jp}\\ 
E-mail: $^\#$ \email{oka(at)ap.t.u-tokyo.ac.jp}\\
E-mail: $^\dagger$ \email{sonoda(at)het.phys.sci.osaka-u.ac.jp}\\ 
}

\abstract{
Using the AdS/CFT correspondence, we calculate the vacuum decay rate 
for the Schwinger effect in confining large $N_{c}$ gauge theories.
The instability is induced by the quark antiquark pair creation
triggered by strong electromagnetic fields. The decay rate is obtained as the imaginary part of the Euler-Heisenberg effective Lagrangian evaluated from the D-brane action with a constant electromagnetic field in holographic QCD models such as the Sakai-Sugimoto model and the deformed Sakai-Sugimoto model. The decay rate is found to increase with the magnetic field parallel to the electric field, while it decreases with the magnetic field perpendicular to the electric field. 
We discuss generic features of a critical electric field as a function of the magnetic field and the QCD string tension in the Sakai-Sugimoto model.
}

\preprint{
{\normalsize OU-HET-842} \\
{\normalsize RIKEN-MP-99}
}


\begin{document}
\setcounter{page}{1}

\section{Introduction}

Schwinger effect is one of the most interesting phenomena in particle physics.
This is a phenomenon that a pair creation of charged particles occur under an external field such as an electromagnetic field. 
Schwinger obtained  the creation rate of an electron positron pair by evaluating the imaginary part of Euler-Heisenberg Lagrangian, which is an effective Lagrangian for a constant electric field \cite{Heisenberg:1935qt,Schwinger:1951nm}.
This rate $\Gamma$ is derived as $\Gamma\sim\mathrm{exp}\left(-\pi m_{e}^{2}/eE\right)$ to leading order and has a form with a negative power in the gauge coupling $e$. So the Scwinger effect is a non-perturbative effect.
Here, $m_{e}$ is  the electron mass and $E$ is an electric field.
A critical electric field necessary energy for the electron positron pair creation is $E_{\mathrm{cr}}\sim m_{e}^{2}c^{3}/e\hbar$, and the strength is about $10^{18}$ $[\mathrm{V}/\mathrm{m}]$.
So, it is a phenomenon which shows up only under strong electromagnetic fields.

Recently, we have seen advance in research on a strong electromagnetic field in both theoretical and experimental aspects of hadron physics.
At the heavy ion collision in RHIC and LHC, it is expected that a strong magnetic field is generated by a collision of charged particles accelerated at about the speed of light. 
Another related topic is neutron stars and magnetors which carry a strong electromagnetic field. In such a strong electromagnetic field, it may be possible to generate a pair creation of charged particles.
For example, we may think of a quark antiquark pair creation as well as the electron positron pair. 
We need to analyze a non-perturbative effect because the quark dynamics is governed by a strongly coupled gauge theory. 
This is becoming possible thanks to a development in calculating physical observables of strongly coupled gauge theories from classical gravity by the AdS/CFT correspondence \cite{Maldacena:1997re,Gubser:1998bc,Witten:1998qj}.

Within the AdS/CFT framework, the quark pair creation rate in the strongly coupled $\mathcal{N}=4$ supersymmetric Yang-Mills theory was obtained in \cite{Semenoff:2011ng,Gorsky:2001up}.
Based on \cite{Semenoff:2011ng,Gorsky:2001up}, the holographic Schwinger effect were calculated in various systems \cite{Ambjorn:2011wz,Bolognesi:2012gr,Sato:2013pxa,Sato:2013iua,Sato:2013dwa,Sato:2013hyw,Kawai:2013xya,Sakaguchi:2014gpa}.
On the other hand, two of the present authors obtained the vacuum decay rate, which can be identified as the creation rate of quark-antiquark pairs, in $\mathcal{N}=2$ supersymmetric QCD(SQCD) by using a different method \cite{Hashimoto:2013mua} in AdS/CFT correspondence: {\it the imaginary part of the probe D-brane action.}\footnote{The method based on \cite{Semenoff:2011ng,Gorsky:2001up} is a single instanton process for the creation of a pair and is valid for the electric field $E$ smaller than the critical electric field, while the method in \cite{Hashimoto:2013mua} is for $E$ stronger than or comparable to the critical electric field. Both are basically a disc partition function in string theory, but evaluated in different regimes. The former is a semi-classical large disc, while the latter is a small disc giving the Dirac-Born-Infeld action. The boundary of the disc corresponds to the world line of the created quark pair.
A small $E$ means a large disc , {\it i.e.} a larger separation of the created quark pair.}
D3-D7 brane system corresponds to $\mathcal{N}=4$ supersymmetric $SU(N_{c})$ Yang-Mills theory including an $\mathcal{N}=2$ hypermultiplet in the fundamental representation of the $SU(N_{c})$ gauge group \cite{Karch:2002sh}.
They obtained the creation rate of the quark antiquark in the $\mathcal{N}=2$ SQCD under a constant electric field by evaluating the imaginary part of the D7-brane action.
Then, the present authors evaluated the imaginary part of the D7-brane action including not only a constant electric field but also a constant magnetic field and obtained the creation rate of the quarks and antiquarks in the $\mathcal{N}=2$ SQCD \cite{Hashimoto:2014dza}.

We summarize the properties of the creation rate in both electric and magnetic 
fields obtained in \cite{Hashimoto:2014dza} for $\mathcal{N}=2$ SQCD as follows.
We derived the Euler-Heisenberg Lagrangian for a constant electromagnetic field in $\mathcal{N}=2$ SQCD at large $N_{c}$ and at strong coupling. 
Then, we obtained the creation rate of the quarks and antiquarks by evaluating the imaginary part of the Lagrangian.
We found that the creation rate diverges at a zero temperature in the massless quark limit while it becomes finite when we introduce a nonzero temperature. 
The divergence of the creation rate is influenced not only by a constant electric field but also by a constant magnetic field.
The results in SQCD showed similarities with the creation rate of the electron positron pair in $\mathcal{N}=2$ supersymmetric QED(SQED) in constant electromagnetic field.

In this paper, we study the quark antiquark pair creation in {\it non-supersymmetric} QCD at large $N_{c}$ at strong coupling, and the imaginary part of D8-brane action in a constant electromagnetic field. The holographic models are the Sakai-Sugimoto model \cite{Sakai:2004cn} and its deformed version \cite{Aharony:2006da}.
Our findings in this paper are as follows:
\begin{itemize}
 \item We derive the Euler-Heisenberg Lagrangian for confining gauge theories: the Sakai-Sugimoto model and the deformed Sakai-Sugimoto model. We obtain the creation rate of the quark antiquark pair under the electromagnetic field, by evaluating the imaginary part of the D-brane actions.
 \item
 The imaginary part is found to increase with the magnetic field parallel to the electric field,
 while it decreases with the magnetic field perpendicular to the electric field. So the vacuum
 instability strongly depends on the direction of the applied magnetic field relative to the electric field.
 \item We obtain a critical value of the electric field, i.e., the Schwinger limit, by using the condition that the D-brane action has the imaginary part. In the case of the Sakai-Sugimoto model, the critical electric field corresponds to a QCD string tension between a quark and an antiquark.
\end{itemize}
As for the first part among above, a result with only an electric field was reported in \cite{Kim:2008zn}. We analyze generic electric and magnetic fields in this paper.

The organization of this paper is as follows. In section 2, we summarize the behavior of the critical electric field under a magnetic field. In section 3, we derive the creation rate of the quark antiquark pair from the imaginary part of the Euler-Heisenberg Lagrangian in the the Sakai-Sugimoto model by using the AdS/CFT correspondence.
Also, in section 4, we consider the imaginary part of the D-brane action in the deformed the Sakai-Sugimoto model. Section 5 is for summary and discussion.

%%%%%%%%%%%%%%%%%%%%%
\section{Universal behavior of the critical electric field}

In this section, 
we derive an expression for the critical electric field $E_{\rm cr}$ in generic holographic QCD beyond which the Euler-Heisenberg Lagrangian acquires an imaginary part
in the presence of a magnetic field. This part follows analyses by Sato and Yoshida
done in \cite{Sato:2013pxa} and \cite{Sato:2013hyw}.
Then, we will find that the expression coincides
with that of QED in the strong magnetic field limit.

First, in any holographic QCD model, it is known that there is an ``IR wall" at which the 
geometry is terminated in the holographic radial direction. 
The renowned Gibbons-Maeda geometry \cite{Gibbons:1987ps,Witten:1998zw} for confining 
pure Yang-Mills dual is one of the best examples. It has the radial scale 
typically written as $U_{\rm KK}$, and the region $U<U_{\rm KK}$ is cut out smoothly
and any physical excitations coming down from the boundary of the spacetime 
should be reflected back at the IR wall. The IR wall is an essential ingredient in 
any bottom-up holographic model for implementing the confining scale.
So the generic confining geometry should have the following form
\begin{eqnarray}
ds^2 = g(r)\eta_{\mu\nu} dx^\mu dx^\nu + f(r) dr^2 + h(r) \mbox{[internal space]}
\label{geombg}
\end{eqnarray}
in which every function of $r$ terminates at some value of $r$ where the IR wall exists.
Here $r$ is the holographic radial coordinate, and $\mu,\nu=0,1,2,3$ is our space time
directions. In the above, the internal space can be generic, and can even mix with the $r$ 
coordinate if one wishes. The important factor is only $g(r)$, as we shall see below.

We consider a flavor D-brane representing the quark sector, and its generic form is
given by the Dirac-Born-Infeld (DBI) action
\begin{eqnarray}
S_{\rm flavor} = -T_{{\rm D}p} \int d^{p+1}\xi \; e^{-\phi}
\sqrt{-\det(\tilde{g}_{ij} + 2\pi\alpha' F_{ij})} \, .
\label{DBI}
\end{eqnarray}
Here $\phi$ is the background dilaton, and $\tilde{g}_{ij}$ is the induced metric on the
D-brane (as the flavor D-brane is curved in the curved background geometry (\ref{geombg})),
\begin{eqnarray}
\tilde{g}_{ij} \equiv g_{MN} \partial_i X^M \partial_j X^N \, .
\end{eqnarray}
$X^M(\xi)$ are the worldvolume scalar fields which specify the position of the flavor D-brane in the bulk spacetime.
The indices $i$ and $j$ run from 0 to $p+1$, the dimension of the worldvolume of the D-brane.
The field strength $F$ can have various components, but our interest is only the 1+3 dimensional
spacetime electromagnetic field which is constant, $\vec{E}$ and $\vec{B}$.
This constant electromagnetic field can satisfy the equations of motion of the DBI 
theory (\ref{DBI}) since everything on the static D-brane is consistently assumed to depend 
only on $r$. So, given constant $\vec{E}$ and $\vec{B}$,  once the scalar field $X(r)$ is solved,
the static D-brane configuration is determined.

Now, we put a simple assumption: the flavor D-brane hits the IR wall. 
The D-brane reaches the bottom of the geometry, which is a natural assumption for 
confining gauge theories with a quark mass less than the QCD dynamical scale.
For example, the Sakai-Sugimoto model \cite{Sakai:2004cn} with the flavor D8-brane placed at the antipodal points
on the Kalza-Klein circle has this property. There are other models sharing this property.
The assumption is necessary to show the critical electric field formula.

Let us calculate the critical electric field. The definition of the critical electric field is the value 
beyond which the effective Euler-Heisenberg (EH) action obtains an imaginary part.
The EH action is nothing but the flavor D-brane action evaluated with the 
constant field strength $F$ \cite{Hashimoto:2013mua}. The DBI action density 
is either real or pure imaginary, so, there exists some $r=r_*$ at which we 
have a vanishing DBI action, in general if the EH has an imaginary part:
\begin{eqnarray}
\sqrt{-\det (\tilde{g}_{\mu\nu}+2\pi\alpha' F_{\mu\nu})} = 0 \, .
\end{eqnarray}
For $E<E_{\rm cr}$, there exist no $r_*$ which satisfies this equation, so there appears
no imaginary part in the Euler-Heisenberg action. However, with $E$
beyond the critical $E_{\rm cr}$, 
there appears some $r=r_*$ on the flavor D-brane and the effective action obtains an imaginary part and becomes unstable. At the critical $E=E_{\rm cr}$, one should find $r_*$ at the IR 
bottom of the D-brane, which is required by a consistency. So one finds 
\begin{eqnarray}
\det \left[
g(r_*) \eta_{\mu\nu} + 2\pi\alpha' F_{\mu\nu}
\right]=0 \, 
\end{eqnarray}
at the critical $E_{\rm cr}$. After a simple calculation one finds
\begin{eqnarray}
E_{\rm cr} = \frac{g(r_*)}{2\pi\alpha'} 
\sqrt{\frac{1 + \left(\frac{2\pi\alpha'}{g(r_*)}\right)^2 |\vec{B}|^2}{1 + \left(\frac{2\pi\alpha'}{g(r_*)}\right)^2 |\vec{B}_{/\!/}|^2}}
\label{EcrB}
\end{eqnarray}
where $\vec{B}_{/\!/}$ is the components of the constant magnetic field which are parallel to the electric field. For the case of ${\cal N}=4$ supersymmetric Yang-Mills theory where the probe brane is a D3-brane, 
this formula was first found in \cite{Sato:2013pxa}. 
We find a generalization of it applicable to any probe and background. 

In particular, when $\vec{B}_\perp$ is zero, we find an expression
\begin{eqnarray}
E_{\rm cr} (\vec{B}_\perp=0) = \frac{g(r_*)}{2\pi\alpha'} \, .
\label{EcrB0}
\end{eqnarray}
Interestingly, this is independent of the parallel magnetic field. This is natural because charged particle moving along the electric field do not feel the Lorentz force if the magnetic field is parallel to the electric field.
Note that this quantity (\ref{EcrB0}) is written only in terms of a single metric component at the IR bottom of the confining geometry.
For a  generic magnetic field, we find an inequality
\begin{eqnarray}
E_{\rm cr}(\vec{B}) \geq E_{\rm cr}(\vec{B}=0)
\end{eqnarray}
which is nothing but the magnetic catalysis.

The particular value of the critical electric field (\ref{EcrB0}) for $\vec{B}_\perp=0$
in fact coincides with the QCD string tension $\sigma_{\rm string}$ \cite{Sato:2013hyw}.
%Now, to show the equality (\ref{eqder}), let us compute the QCD string tension.
The QCD string tension is just the effective tension of a fundamental string at the
bottom of the geometry. The string worldsheet should be along the time direction and some
spatial direction $\mu=1,2,3$, as being consistent with the Regge behavior, so
\begin{eqnarray}
\sigma_{\rm string} = T_{\rm F1} \sqrt{-g_{00} g_{11}}\biggm|_{\rm IR \; bottom} \, .
\end{eqnarray}
Substituting the background metric (\ref{geombg}) and using the fundamental string tension 
$T_{\rm F1}=1/2\pi\alpha'$, we obtain
\begin{eqnarray}
\sigma_{\rm string} = \frac{1}{2\pi\alpha'} \sqrt{-g(r_*)\eta_{00} g(r_*)\eta_{11}}
=\frac{g(r_*)}{2\pi\alpha'}\, .
\label{QCDst}
\end{eqnarray}
The value again is written solely by a single component of the metric at the IR bottom,
and coincides completely with the critical electric field
(\ref{EcrB0}). %Therefore, we have verified the equality (\ref{eqder}).

%In the presence of the magnetic field, the equality (\ref{eqder}) does not hold. It gives
%the minimum electric field to produce the instability, and the magnetic field generically
%pushes the minimum up. 
Using the relation (\ref{QCDst}), from (\ref{EcrB}) we find a formula for the critical electric field
in the presence of the generic magnetic field as 
%and we can conjecture that this would be a universal feature of the
%critical magnetic field,
\begin{eqnarray}
E_{\rm cr} = \sigma_{\rm string} 
\sqrt{\frac{\sigma_{\rm string}^2 + |\vec{B}|^2}{\sigma_{\rm string}^2 + |\vec{B}_{/\!/}|^2}} \, .
\label{EcrBst}
\end{eqnarray}
The critical electric field is shown in Fig.~\ref{fig:E}. One can see that the magnetic field perpendicular to the electric field makes the critical electric field to increase.
%%%%%%%%%%%%%%%%%%%%%%%
\FIGURE[t] {
\includegraphics[width=7cm]{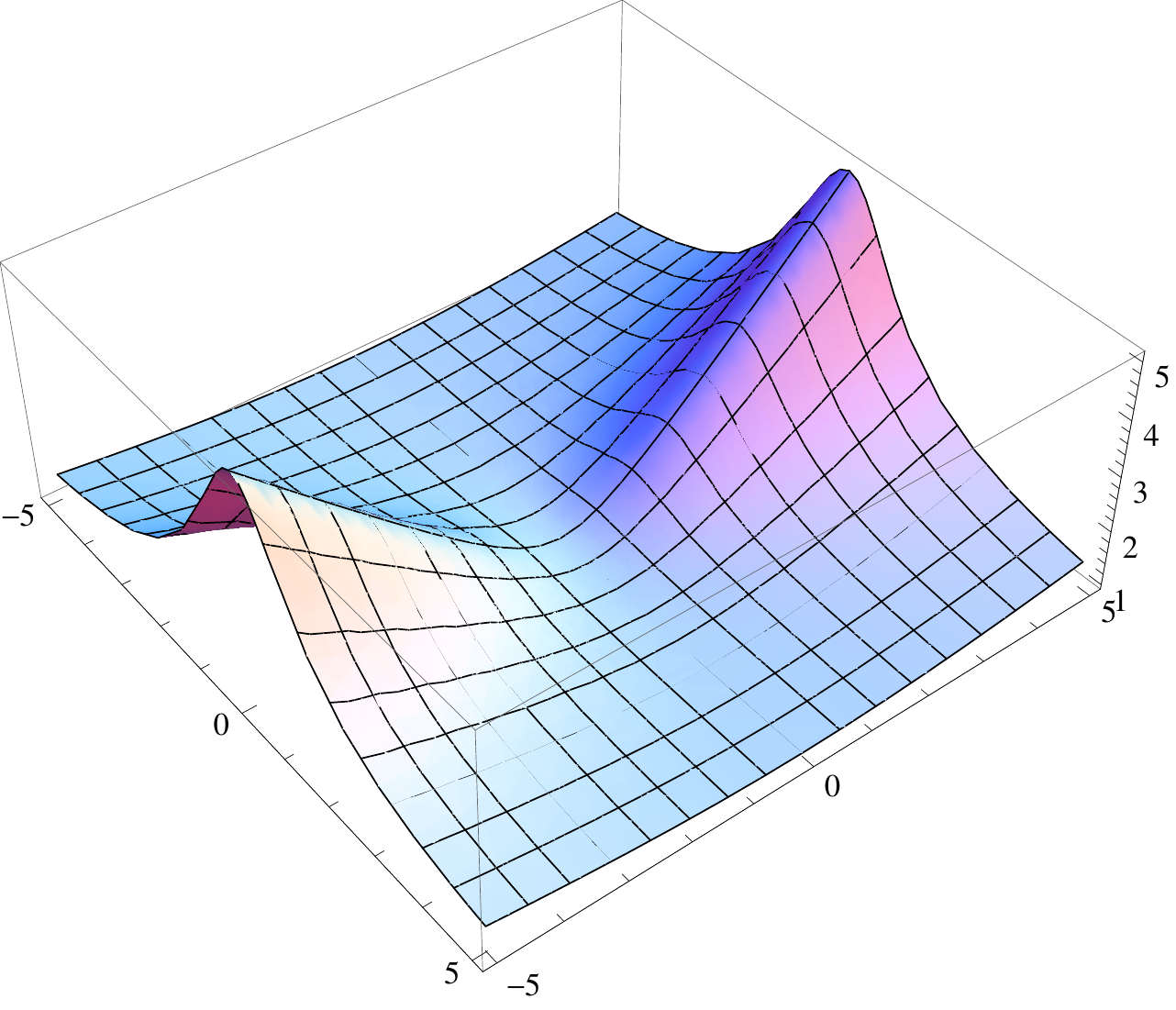}
\put(-160,20){$B_{\!/\!/}$}
\put(-70,20){$B_{\perp}$}
\put(0,100){$E_{\rm cr}$}
\caption{The critical electric field formula $E_{\rm cr}$ as a function of the magnetic field. 
$B_{\!/\!/}$ is the magnetic field parallel to the electric field, and $B_{\perp}$ is the one perpendicular to the electric field.
}
\label{fig:E}
}
%%%%%%%%%%%%%%%%%%%%%%%
When there is no magnetic field, this formula reduces to
\begin{eqnarray}
E_{\rm cr} = \sigma_{\rm string}
\label{eqder}
\end{eqnarray}
which states that the critical electric field coincides with the confining force 
(the QCD string tension) $\sigma_{string}$ between a quark and an antiquark.
The equality is quite naturally interpreted in QCD. Quarks are charged under the electric field,
while the quark is bound to an antiquark with a confining force. So, if the external electric field
is stronger than the confining force, the quarks are liberated, and electric current would start to flow. It is a phase transition to a non-equilibrium steady state, and naively the critical electric field is expected to be equal to the confining force, that is, the QCD string tension. 

Remember that we have assumed that the flavor D-brane hits the IR wall, to derive this equation. For some AdS/CFT models, once the magnetic field is turned on, the assumption may not be satisfied (for example, see \cite{Filev:2007gb,Evans:2011mu,Filev:2010pm}). 
So the above formula applies only a sub-class of the AdS/CFT models.

The formula shows 
that in the presence of the magnetic field, the critical electric field has a particular dependence on
the magnetic field. It is interesting to note that the dependence agrees with what is expected in QED 
in strong magnetic field. In QED, electrons form Landau levels in the magnetic field,
and for a strong magnetic field only the lowest Landau level is expected to contribute to the
dynamics. The lowest Landau level approximation provides 
a decay rate of the vacuum in the strong magnetic field as \cite{Hidaka:2011fa}
\begin{eqnarray}
{\rm Im}\, {\cal L}_{\rm QED} 
\sim \frac{{\cal E}{\cal B}}{4\pi^2} \log \left[\frac{1}{1-\exp[-\pi m^2/{\cal E}]}\right] \, ,
\label{QEDL}
\end{eqnarray}
where 
\begin{eqnarray}
{\cal E} \equiv \sqrt{\sqrt{F^2+G^2}-F}, \quad
{\cal B} \equiv \sqrt{\sqrt{F^2+G^2}+F}, \quad
\end{eqnarray}
with the Lorentz invariant combinations of the electromagnetic field,
\begin{eqnarray}
F \equiv (\vec{B}^2-\vec{E}^2)/2, \quad G \equiv \vec{B}\cdot\vec{E}\, .
\end{eqnarray}
The exponent appearing in (\ref{QEDL}) is expected to be corrected 
\cite{Affleck:1981bma} at a finite coupling constant
$e$ as $-\pi m^2/{\cal E} \rightarrow -\pi m^2/{\cal E} + e^2/4$.
The QED expression (\ref{QEDL}) will become singular if the exponent vanishs,
which occurs at a certain value of ${\cal E}$,
\begin{eqnarray}
{\cal E} = \tilde{\sigma}
\end{eqnarray}
with $\tilde{\sigma}=\frac{4\pi m^2}{e^2}$.
Using the definition of ${\cal E}$, this equation is solved as
\begin{eqnarray}
E_{\rm cr} = \tilde{\sigma}
\sqrt{\frac{\tilde{\sigma}^2 + |\vec{B}|^2}{\tilde{\sigma}^2 + |\vec{B}_{/\!/}|^2}} \, .
\end{eqnarray}
This expression is exactly the same as what we found in the D-brane analysis, (\ref{EcrBst}).
It is intriguing that our generic formula derived from string theory with the DBI action coincides with
the QED expectation at strong magnetic field. 

%%%%%%%%%%%%%%%%%%%%%%%%%%%%%%%%%%%

\section{Pair creation of quark antiquark in D4-D8 brane system}
In this section, we study a quark antiquark pair creation in the confining phase. The Sakai-Sugimoto model is the D-brane construction of the D4-D8 brane which has the $SU(N_{f})_{L}\times SU(N_{f})_{R}$ chiral symmetry and the confining phase \cite{Sakai:2004cn}.
We will obtain the creation rate of the quark antiquark in the confining non-supersymmetric gauge theory by evaluating the imaginary part of the D8-brane action with a constant electromagnetic field.
Also, the critical electric field is obtained by a threshold at which the D8-brane action acquires a non-vanishing imaginary part.
\subsection{Review of the Sakai-Sugimoto model }
The D-brane contruction of the Sakai-Sugimoto model is with $N_{c}$ D4- and D8-branes. 
A spatial coordinate $x^{4}$ of the spatial world-volume directions is compactified on $S^{1}$ with an anti-periodic boundary conditions for the fermions. The $N_{f}$ D8-branes intersect $x^{4}=0$ with the D4-branes. Similarly, the $N_{c}$ anti-D8-branes put parallel at $x^{4}=\pi R$.
Here, the $R$ is the radius of $S^{1}$. We consider a flavor $N_{f}=1$ for simplicity in this paper.

The D4-branes metric is
\begin{align}
ds_{D4}^2 = \left(\frac{u}{R_{D4}}\right)^{3/2}(-dt^2 + \delta_{ij}dx^{i}dx^{j} + f(u)(dx^{4})^{2}) + \left(\frac{R_{D4}}{u}\right)^{3/2}\left(\frac{du^{2}}{f(u)} + u^{2}d\Omega_{4}^{2}\right). \label{D4BRANE}
\end{align}
The dilaton, the field strength of the Ramond-Ramond field, the function $f(u)$ and the AdS radius are defined as follows,
\begin{align}
e^{\phi} = g_{s}\left(\frac{u}{R_{D4}}\right)^{3/4}, \hspace{3mm} F_{4}\equiv dC_{3} = \frac{2\pi N_{c}}{V_{4}}\epsilon_{4}, \hspace{3mm} f(u)\equiv 1- \frac{u_{KK}^{3}}{u^{3}},\hspace{3mm}R^{3}_{D4}\equiv \pi g_{s}N_{c}l_{s}^{3} ,\label{dic}
\end{align}
where $g_{s}$ is a string coupling and $N_{c}$ is the number of colors  gauge group.  String length is $l_{s}$ and is related to $\alpha'$ as $l^{2}_{s}=\alpha'$. The coordinate $u$ is the holographic radial direction, and $u=\infty$ corresponds to the boundary of the bulk space. The coordinate $u$ is defined for the region $u_{KK}\le u\le\infty$.
$V_{4}$ is the volume of the unit four sphere $S^{4}$. $\epsilon_{4}$ is the volume form of the $S^{4}$. In order to avoid a possible singularity at $u=u_{KK}$, the coordinate $u$ is follows a periodic boundary condition as follows,
\begin{align}
x^{4}\sim x^{4} + \delta x^{4}, \hspace{4mm} \delta x^{4}\equiv \frac{4\pi}{3}\frac{R_{D4}^{3/2}}{u_{KK}^{1/2}}=2\pi R.
\end{align}
The Kaluza-Klein mass parameter is defined as follows,
\begin{align}
M_{KK} \equiv \frac{2\pi}{\delta x^{4}} = \frac{3}{2}\frac{u^{1/2}_{KK}}{R^{3/2}_{D4}}.
\end{align}
The gauge coupling $g_{YM}$ at the cutoff scale $M_{KK}$ in the 4-dimensional Yang-Mills theory is derived as $g_{YM}^{2}=(2\pi)^{2}g_{s}l_{s}/\delta x^{4}$ from the D4-brane action compactified on $S^{1}$. Thus, the AdS/CFT dictionary which is the relationship between the parameters $R_{D4}, u_{KK}, g_{s}$ in the gravity side and the parameters $M_{KK}, g_{YM}$'Æ$N_{c}$ in the gauge side is the following,
\begin{align}
R^{3}_{D4} = \frac{1}{2}\frac{\lambda l_{s}^{2}}{M_{KK}}, \hspace{5mm} u_{KK} = \frac{2}{9}\lambda M_{KK}l_{s}^{2}, \hspace{5mm} g_{s} = \frac{1}{2\pi}\frac{\lambda}{M_{KK}N_{c}l_{s}},
\end{align}
where a 't Hooft coupling $\lambda$ is defined as $\lambda\equiv g_{YM}^{2}N_{c}$.

Next, we consider a D8-brane embedded in the D4-brane background. The D8-brane and the anti-D8-brane are inserted respectively to $x^{4}=0$ and $x^{4}=\pi R$.  Under this boundary condition, the equation of motion requires $dx^{4}/du=0$ which means that the coordinate $x^{4}$ of the D8-brane and anti-D8-brane is constant. Then, the induced metric on the D8-brane is
\begin{align}
ds_{D8}^2 = \left(\frac{u}{R_{D4}}\right)^{3/2}(-dt^2 + \delta_{ij}dx^{i}dx^{j}) + \left(\frac{R_{D4}}{u}\right)^{3/2}\left(\frac{du^{2}}{f(u)} + u^{2}d\Omega_{4}^{2}\right).
\end{align}
The D8-brane action  is represented by
\begin{align}
S_{D8} = S_{D8}^{\mathrm{DBI}} + S_{D8}^{\mathrm{CS}}.
\end{align}
The $S^{\mathrm{DBI}}_{D8}$ is the D8-brane Dirac-Born-Infeld(DBI) action and the $S^{\mathrm{CS}}_{D8}$ is the D8-brane Chern-Simons term. 
We do not consider the Chern-Simons term in this paper.

\subsection{Euler-Heisenberg Lagrangian of the Sakai-Sugimoto model}

We shall calculate the Euler-Heisenberg Lagrangian. It is simply the DBI action with a constant electromagnetic field.
We substitute the D8-brane background and a constant electromagnetic field to the DBI action. The constant electromagnetic field on the $S^{4}$ is zero. We turn on only the electric field on the $x^{1}$ direction without losing generality due the spacial rotational symmetry. The magnetic fields are introduced in $x^{1},x^{2},x^{3}$ directions. The DBI action in the D8-brane background including a constant electromagnetic field is given by
\begin{align}
S_{D8}^{\mathrm{DBI}} = - T_{8}\int{d^{4}xdud\Omega_{4}}e^{-\phi}\sqrt{-\mathrm{det}(P[g]_{ab} + 2\pi\alpha'F_{ab})}\label{D8DBI},
\end{align}
where $T_{8}$ is a D8-brane tension and defined as $T_{8}=1/(2\pi)^{8}l_{s}^{9}$. Substituting the D8-brane background and the constant electromagnetic field to the D8-brane action, the effective Lagrangian is obtained by
\begin{align}
\mathcal{L} = - \frac{8\pi^{2}}{3}T_{8}\int_{u_{KK}}^{\infty}{du}\hspace{1mm}e^{-\phi}\frac{u^{4}}{\sqrt{f(u)}}\left(\frac{R_{D4}}{u}\right)^{3/4}\sqrt{\xi},
\end{align}
where the $d\Omega_{4}$ integral is Vol($S^{4}$)$=$$8\pi^{2}/3$. Here $\xi$ is defined by
\begin{align}
\xi&\equiv 1 - \frac{(2\pi\alpha')^{2}R_{D4}^{3}}{u^{3}}\left[F_{01}^{2} - F_{12}^{2} - F_{23}^{2} - F_{13}^{2} + f(u)\frac{u^{3}}{R_{D4}^{3}}(F_{0u}^{2} - F_{1u}^{2})\right] \notag \\
   & -  \frac{(2\pi\alpha')^{4}R_{D4}^{6}}{u^{6}}\left[F_{01}^{2}F_{23}^{2} + f(u)\frac{u^{3}}{R_{D4}^{3}}\{F_{0u}^{2}(F_{12}^{2} + F_{23}^{2} + F_{13}^{2}) - F_{1u}^{2}F_{23}^{2}\} \right].
\end{align}

Next, we derive the equations of motion from the DBI action. We put $\partial_{i}=0,\hspace{1mm}(i=1,2,3)$ because we are interested in homogeneous phases. The equations of motion are given by \footnote{When both the electric and the magnetic fields are nonzero, the Chern-Simons term
comes into the equations of motion. Since the Chern-Simons term is of the form $\sim A_u E B$, the equations of motion for $A_u$ acquires a new term, $\partial_0 A_u \sim EB$. This is nothing but the
chiral anomaly. The field $A_u$ grows in time for a constant $E$ and $B$. We ignore this anomaly 
effect for simplicity, 
and interpret our outcome as the physical values measured at $t=0$ at which $A_u$ vanishes
as an initial condition.}
\begin{align}
\frac{(2\pi\alpha')^{2}8\pi^{2}T_{8}}{3g_{s}}\partial_{u}\left[\frac{(R_{D4}/u)^{3/2}u^{4}\sqrt{f(u)}F_{0u}\left(1 + \frac{(2\pi\alpha')^{2}R_{D4}^{3}}{u^{3}}\right)(F_{12}^{2} + F_{23}^{2} + F_{13}^{2})}{\sqrt{\xi}}\right] = 0,  \\ \notag
\end{align}
\begin{align}
\frac{(2\pi\alpha')^{2}8\pi^{2}T_{8}}{3g_{s}}\partial_{0}\left[\frac{(R_{D4}/u)^{3/2}u^{4}\sqrt{f(u)}F_{0u}\left(1 + \frac{(2\pi\alpha')^{2}R_{D4}^{3}}{u^{3}}\right)(F_{12}^{2} + F_{23}^{2} + F_{13}^{2})}{\sqrt{\xi}}\right] = 0, \\ \notag
\end{align}
\begin{align}
\frac{(2\pi\alpha')^{2}8\pi^{2}T_{8}}{3g_{s}}\partial_{0}\left[\frac{(R_{D4}/u)^{9/2}u^{4}F_{01}\left(1 + \frac{(2\pi\alpha')^{2}R_{D4}^{3}}{u^{3}}F_{23}^{2}\right)}{\sqrt{\xi f(u)}}\right] + \notag \\ 
\frac{(2\pi\alpha')^{2}8\pi^{2}T_{8}}{3g_{s}}\partial_{u}\left[\frac{(R_{D4}/u)^{3/2}u^{4}\sqrt{f(u)}F_{1u}\left(1 + \frac{(2\pi\alpha')^{2}R_{D4}^{3}}{u^{3}}F_{23}^{2}\right)}{\sqrt{\xi}}\right] = 0. 
\end{align}
In particular, the equations of motion for static configurations are derived as
\begin{align}
\frac{(2\pi\alpha')^{2}8\pi^{2}T_{8}}{3g_{s}}\partial_{u}\left[\frac{(R_{D4}/u)^{3/2}u^{4}\sqrt{f(u)}F_{0u}\left(1 + \frac{(2\pi\alpha')^{2}R_{D4}^{3}}{u^{3}}\right)(F_{12}^{2} + F_{23}^{2} + F_{13}^{2})}{\sqrt{\xi}}\right] = 0,
\end{align}
\begin{align}
\frac{(2\pi\alpha')^{2}8\pi^{2}T_{8}}{3g_{s}}\partial_{u}\left[\frac{(R_{D4}/u)^{3/2}u^{4}\sqrt{f(u)}F_{1u}\left(1 + \frac{(2\pi\alpha')^{2}R_{D4}^{3}}{u^{3}}F_{23}^{2}\right)}{\sqrt{\xi}}\right] = 0.
\end{align}
By using the equations of motion, we can derive the charge density $d$ and the current density $j$ respectively as,
\begin{align}
d &\equiv \frac{(2\pi\alpha')^{2}8\pi^{2}T_{8}}{3g_{s}}\frac{(R_{D4}/u)^{3/2}u^{4}\sqrt{f(u)}F_{0u}\left(1 + \frac{(2\pi\alpha')^{2}R_{D4}^{3}}{u^{3}}\right)(F_{12}^{2} + F_{23}^{2} + F_{13}^{2})}{\sqrt{\xi}}, \\
j &\equiv \frac{(2\pi\alpha')^{2}8\pi^{2}T_{8}}{3g_{s}}\frac{(R_{D4}/u)^{3/2}u^{4}\sqrt{f(u)}F_{1u}\left(1 + \frac{(2\pi\alpha')^{2}R_{D4}^{3}}{u^{3}}F_{23}^{2}\right)}{\sqrt{\xi}}.
\end{align}
In this paper, we are not interested in the charge density and the current as we are looking at the vacuum instability. So we put $F_{0u}=0$ and $F_{1u}=0$ consistently.

Therefore, the D8-brane Lagrangian is derived as
\begin{align}
\mathcal{L} =  - \frac{8\pi^{2}T_{8}}{3g_{s}}\int^{\infty}_{u_{KK}}du\hspace{1mm}\frac{u^{4}(R_{D4}/u)^{3/2}}{\sqrt{1 - \frac{u_{KK}^{3}}{u^{3}}}}\sqrt{1 - \frac{(2\pi\alpha')^{2}R_{D4}^{3}}{u^{3}}\left[E_{1}^{2} - \vec{B}^{2}\right] -  \frac{(2\pi\alpha')^{4}R_{D4}^{6}}{u^{6}}E_{1}^{2}B_{1}^{2}},\label{ELAdS}
\end{align}
where we define the constant electric field as $F_{01}\equiv E_{1}$ and the constant magnetic fields as $F_{12}\equiv B_{3}, F_{23}\equiv B_{1}, F_{13}\equiv B_{2}$, $\vec{B}^{2}\equiv B_{1}^{2} + B_{2}^{2} + B_{3}^{2}$. We change the variable $u$ in this integral to a new coordinate $y$ defined by $u=u_{KK}/y$. By using the dictionary of the AdS/CFT correspondence, we reach the non-supersymmetric Euler-Heisenberg Lagrangian at large $N_{c}$,
\begin{align}
\mathcal{L} = - \frac{M_{KK}^{4}\lambda^{3}N_{c}}{2\cdot 3^{8}\pi^{5}}\int^{1}_{0}{dy}\frac{\sqrt{1 - \frac{3^{6}\pi^{2}}{4M_{KK}^{4}\lambda^{2}}y^{3}(E_{1}^{2} - \vec{B}^{2}) - \left(\frac{3^{6}\pi^{2}}{4M_{KK}^{4}\lambda^{2}}\right)^{2}y^{6}E_{1}^{2}B_{1}^{2}}}{y^{9/2}\sqrt{1 - y^{3}}}.
\end{align}
\subsection{Imaginary part of the effective action in Sakai-Sugimoto model}
In the previous subsection, we obtained the Euler-Heisenberg Lagrangian (\ref{ELAdS}). Let us evaluate the imaginary part from the effective Lagrangian.

We look at the region of the $u$ where the imaginary part of the Euler-Heisenberg Lagrangian (\ref{ELAdS}) appears: the square root of the numerator in the integrand of (\ref{ELAdS}) needs less than zero,
\begin{align}
1 - \frac{(2\pi\alpha')^{2}R_{D4}^{3}}{u^{3}}\left[E_{1}^{2} - \vec{B}^{2}\right] - \frac{(2\pi\alpha')^{4}R_{D4}^{6}}{u^{6}}E_{1}^{2}B_{1}^{2} < 0.\label{CONDITION}
\end{align}
Note that the region of the original integral in (\ref{ELAdS}) is from $u_{KK}$ to $\infty$. The condition for the variable $u$ such that the imaginary part of the Euler-Heisenberg Lagrangian is nonzero is given by
\begin{align}
u_{KK}\leq u \leq \left[\frac{(2\pi\alpha')^{2}R_{D4}^{3}}{2}\left\{ E^{2}_{1} - \vec{B}^{2} + \sqrt{(E^{2}_{1} - \vec{B}^{2})^{2} + 4E_{1}^{2}B_{1}^{2}}\right\} \right]^{1/3}. \label{CESAKAI}
\end{align}
Thus, the imaginary part of the effective Lagrangian is obtained as 
\begin{align}
\mathrm{Im}\mathcal{L} = \frac{8\pi^{2}T_{8}}{3g_{s}}\int^{u_{*}}_{u_{KK}}du\hspace{1mm}\frac{u^{4}(R_{D4}/u)^{3/2}}{\sqrt{1 - \frac{u_{KK}^{3}}{u^{3}}}}\sqrt{\frac{(2\pi\alpha')^{4}R_{D4}^{6}}{u^{6}}E_{1}^{2}B_{1}^{2} + \frac{(2\pi\alpha')^{2}R_{D4}^{3}}{u^{3}}\left[E_{1}^{2} - \vec{B}^{2}\right] - 1},\label{ELAdSIm}
\end{align}
where the $u_{*}$ is defined by
\begin{align}
u_{*} \equiv \left[\frac{(2\pi\alpha')^{2}R_{D4}^{3}}{2}\left\{ E^{2}_{1} - \vec{B}^{2} + \sqrt{(E^{2}_{1} - \vec{B}^{2})^{2} + 4E_{1}^{2}B_{1}^{2}}\right\} \right]^{1/3}. \label{U}
\end{align}
The region of the integral \label{ELAdSIm} is shown in Fig.\ref{fig:SS.pdf}.

In terms of the integral variable $y$, the imaginary part is
\begin{align}
\mathrm{Im}\mathcal{L}=\frac{N_{c}\lambda^{3}M_{KK}^{4}}{2\cdot 3^{8}\pi^{5}}\int^{1}_{y_{*}}dy\frac{\sqrt{\left(\frac{3^{6}\pi^{2}}{4\lambda^{2}M_{KK}^{2}}\right)^{2}y^{6}E^{2}_{1}B^{2}_{1}+\frac{3^{6}\pi^{2}}{4\lambda^{2}M_{KK}^{2}}y^{3}\left[E_{1}^{2} - \vec{B}^{2} \right] -1}}{y^{9/2}\sqrt{1-y^{3}}},
\label{final-ImL}
\end{align}
where $y_{*}$ is defined by
\begin{align}
y_{*} \equiv \left[\frac{3^{6}\pi^{2}}{2^{3}\lambda^{2}M_{KK}^{4}}\left\{ E^{2}_{1} - \vec{B}^{2} + \sqrt{(E^{2}_{1} - \vec{B}^{2})^{2} + 4E_{1}^{2}B_{1}^{2}} \right\} \right]^{-1/3}.
\end{align}

Let us examine whether or not this creation rate of the quark antiquark diverges. We evaluate (\ref{ELAdSIm}) by the neighborhood of $u_{KK}$. When we expand $u=u_{KK}+\epsilon \hspace{1mm}(\epsilon\ll u_{KK})$, the creation rate of the quark antiquark is
\begin{align}
\mathrm{Im}\mathcal{L} &\simeq \frac{8\pi^{2}T_{8}R_{D4}^{3/2}}{3g_{s}}F(u_{KK})\int^{u_{*} - u_{KK}}_{0}d\epsilon\frac{1}{\sqrt{(u_{KK} + \epsilon)^{3}-u_{KK}^{3}}} \notag \\
& \simeq \frac{8\pi^{2}T_{8}R_{D4}^{3/2}F(u_{KK})}{3\sqrt{3}g_{s}u_{KK}}\int^{u_{*} - u_{KK}}_{0}d\epsilon\frac{1}{\sqrt{\epsilon}} = (\mathrm{finite}),
\end{align}
where the function $F(u)$ is defined by
\begin{align}
F(u)\equiv u^{4}\sqrt{\frac{(2\pi\alpha')^{4}R_{D4}^{6}}{u^{6}}E_{1}^{2}B_{1}^{2} + \frac{(2\pi\alpha')^{2}R_{D4}^{3}}{u^{3}}\left[E_{1}^{2} - \vec{B}^{2}\right] - 1}.
\end{align}
In the case of $\epsilon\ll u_{KK}$, we may approximate $F(u_{KK}+\epsilon)\simeq F(u_{KK})$ since it is not divergent. So, the creation rate does not diverge in the Sakai-Sugimoto model. Obviously, this is due to the confining scale $u_{KK}$. 

%%%%%%%%%%%%%%%%%%%%%%%
\FIGURE[t]{ 
\includegraphics[width=4.5cm]{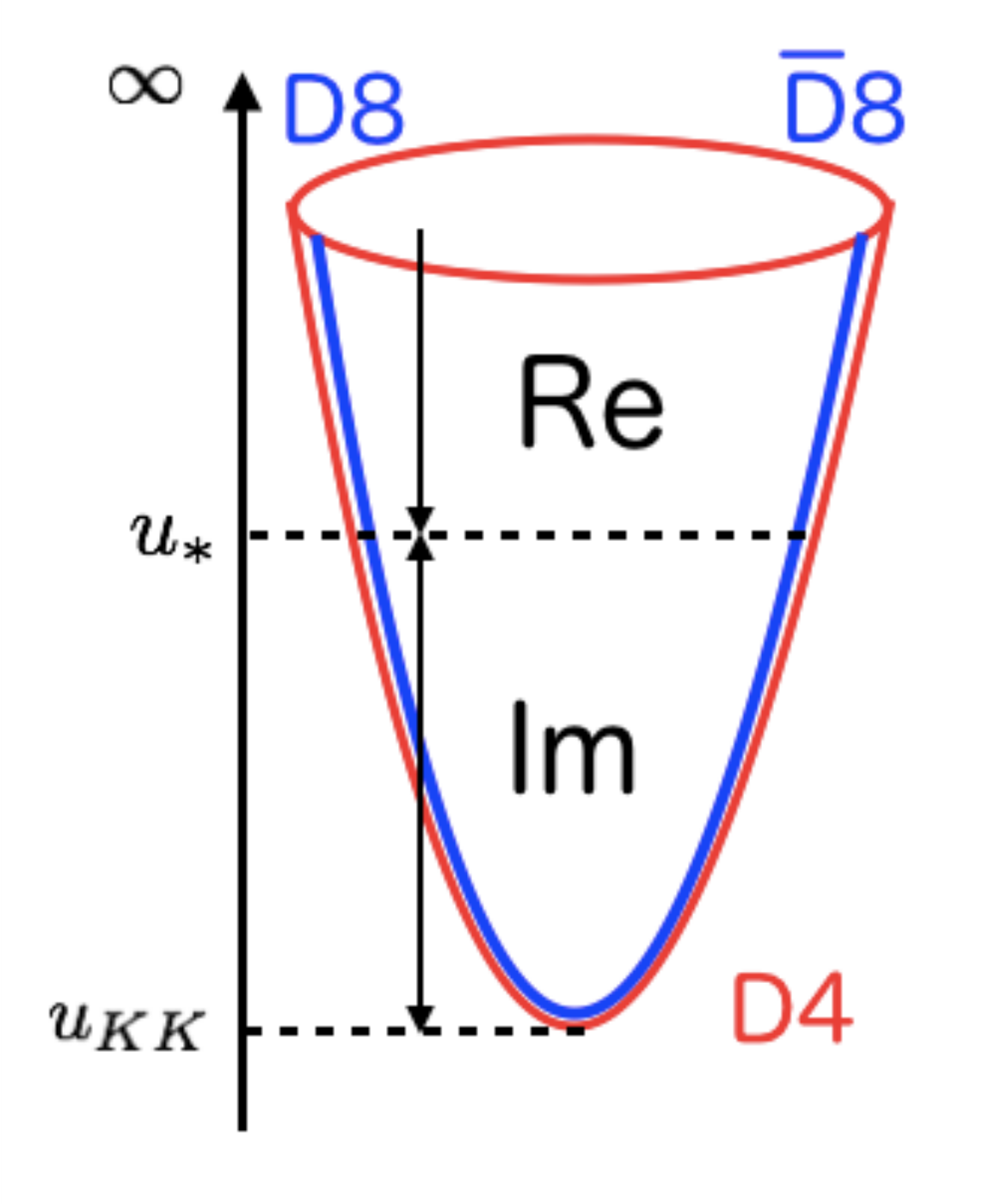}
\caption{
When the region of $u$ is $u_{KK}\le u\le u_{*}$, the Euler-Heisenberg Lagrangian has an imaginary part. It means that 
the pair creation of the quark antiquark occurs by the vacuum instability. 
}
\label{fig:SS.pdf}
}
%%%%%%%%%%%%%%%%%%%%%%%

We evaluate the critical electric field to break the vacuum by the creation of the quark antiquark. We derive the critical electric field from the condition that the effective Lagrangian starts to have the imaginary part. That is, from (\ref{CESAKAI}) we obtain
\begin{align}
u_{KK} \!\leq\! \left[\!\frac{(2\pi\alpha')^{2}R_{D4}^{3}}{2}\!\left\{ \! E^{2}_{1} \!-\! \vec{B}^{2} + \sqrt{(E^{2}_{1} \!-\! \vec{B}^{2})^{2} \!+\! 4E_{1}^{2}B_{1}^{2}}\!\right\} \!\right]^{1/3}.
\end{align}
Thus, the critical electric field $E_{\mathrm{cr}}$ is
\begin{align}
E_{\mathrm{cr}} = \left[\frac{u^{3}_{KK}}{(2\pi\alpha')^{2}R_{D4}^{3}}\cdot\frac{\left\{\frac{u^{3}_{KK}}{(2\pi\alpha')^{2}R_{D4}^{3}} + \vec{B}^{2}\right\}}{\left\{\frac{u^{3}_{KK}}{(2\pi\alpha')^{2}R_{D4}^{3}} + B_{1}^{2}\right\} } \right]^{1/2}. \label{CRESAKAI}
\end{align}
As we can see from (\ref{CRESAKAI}), for $B_{2}=B_{3}=0$, the critical electric field is $E_{\mathrm{cr}} = \left[u_{KK}^{3}/(2\pi\alpha')^{2}R_{D4}^{3}\right]^{1/2}$ and does not depend on $B_{1}$.
By using the dictionary of the AdS/CFT correspondence, the critical electric field is obtained as
\begin{align}
E_{\mathrm{cr}} = \frac{2}{27\pi}\lambda M_{KK}^{2}\left[\frac{\frac{4}{3^{6}\pi^{2}}\lambda^{2}M_{KK}^{4} + \vec{B}^{2}}{\frac{4}{3^{6}\pi^{2}}\lambda^{2}M_{KK}^{4} + B_{1}^{2}}\right]^{1/2}.
\end{align}
This expression coincides with the generic formula (\ref{EcrBst}), since the
QCD string tension of the Sakai-Sugimoto model is $(2/27)\lambda M_{\rm KK}^2$.
When $B_{2},B_{3}=0$, the critical electric field is $E_{\mathrm{cr}}=2\lambda M_{KK}^{2}/27\pi$.

Let us evaluate the imaginary part of the Lagrangian (\ref{final-ImL}).
For a given electric field, the magnetic field can be decomposed into the parallel component and
the perpendicular component. For numerical simplicity, we choose to measure the electric and 
magnetic fields in the unit of $2 \lambda M_{\rm KK}^2/(3^3 \pi)$ and denote those rescaled 
electromagnetic fields as $\tilde{E}$ and $\tilde{B}$. 
Our result (\ref{final-ImL}) is written as
\begin{align}
\mathrm{Im}\mathcal{L}
=\frac{N_{c}\lambda^{3}M_{KK}^{4}}{2\cdot 3^{8}\pi^{5}}\int^{1}_{y_*}dy
\frac{\sqrt{y^6 \tilde{E}^2 \tilde{B}^2_{/\!/} + y^3\left(\tilde{E}^2-\tilde{B}^2_\perp-\tilde{B}^2_{/\!/}\right) -1}}{y^{9/2}\sqrt{1-y^{3}}}
\, .
\label{final-ImL-new}
\end{align}
This can be numerically evaluated, and the result is shown in Fig.\ref{fig:BB}. 
For a fixed electric field, we plot ${\rm Im} {\cal L}$
as a function of the parallel magnetic field $B_{\!/\!/}$ and the perpendicular magnetic field $B_{\perp}$.
%%%%%%%%%%%%%%%%%%%%%%%
\FIGURE[t]{ 
\includegraphics[width=8cm]{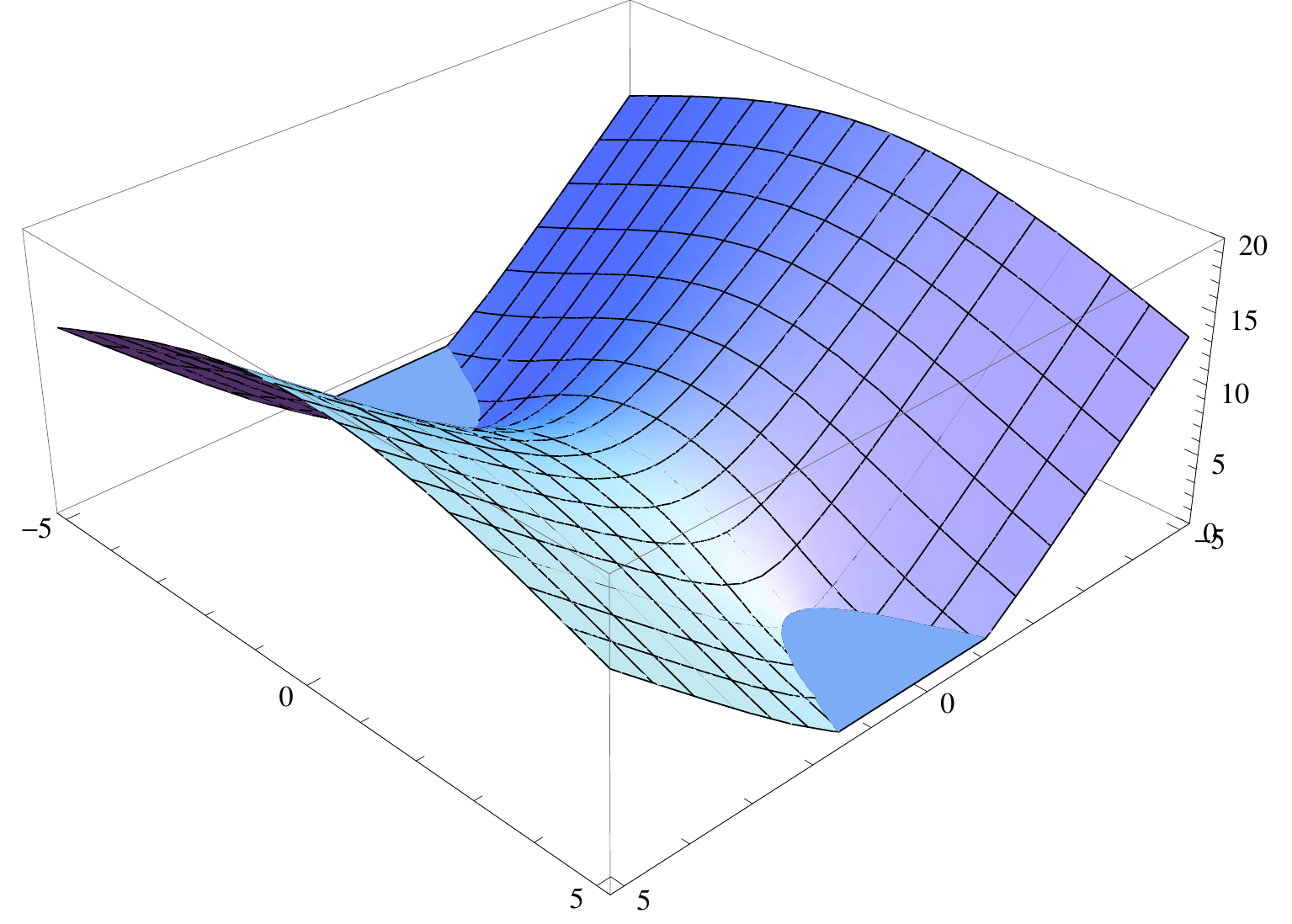}
\put(-180,20){$B_{\perp}$}
\put(-70,20){$B_{\!/\!/}$}
\put(-20,130){Im${\cal L}$}
\caption{The plot of the imaginary part of the Lagrangian for a fixed $E$, as a function of
the magnetic field $B_{\!/\!/}$ parallel to the electric field, and the magnetic field $B_{\perp}$ perpendicular to the electric field. For a large $|B_{\!/\!/}|$, the imaginary
part disappears. We took $\tilde{E}=10$ in this figure.
}
\label{fig:BB}
}
%%%%%%%%%%%%%%%%%%%%%%%
We find that the imaginary part ${\rm Im} {\cal L}$ has 
a very different dependence on these parallel / perpendicular components of the magnetic field.
When the magnetic field is parallel to the electric field, the imaginary part of 
the Lagrangian increases as the parallel magnetic field increases.
On the other hand, when the magnetic field is perpendicular to the electric field,
the situation is completely different. The evaluated imaginary part of the Lagrangian decreases when the perpendicular magnetic field increases.
So, we conclude that the instability of the system is enhanced with the
parallel magnetic field while is suppressed with the perpendicular magnetic field.

The creation rate of the quark antiquark pair is expected to increase with the parallel magnetic field because the magnetic field makes the (1+3)-dimensional system reduce effectively to a (1+1)-dimensional system by a Landau-level quantization. Our result is similar to \cite{Hashimoto:2014dza} in SQCD.

Let us look more about the electric field dependence.
For a parallel
magnetic field, (\ref{final-ImL-new}) is written as
\begin{align}
\mathrm{Im}\mathcal{L}_{\rm para.\;B}
=\frac{N_{c}\lambda^{3}M_{KK}^{4}}{2\cdot 3^{8}\pi^{5}}\int^{1}_{\tilde{E}^{-2/3}}dy
\frac{\sqrt{(y^{3}\tilde{E}^{2}-1)(y^3 \tilde{B}^{2}_{/\!/}+1)}}{y^{9/2}\sqrt{1-y^{3}}}
\, .
\label{final-ImL1}
\end{align}
For a perpendicular magnetic field, it is written as
\begin{align}
\mathrm{Im}\mathcal{L}_{\rm perp.\;B}
=\frac{N_{c}\lambda^{3}M_{KK}^{4}}{2\cdot 3^{8}\pi^{5}}\int^{1}_{(\tilde{E}^2-\tilde{B}^2_\perp)^{-1/3}}dy
\frac{\sqrt{y^{3}(\tilde{E}^{2}-\tilde{B}^2_\perp)+1}}{y^{9/2}\sqrt{1-y^{3}}}
\, .
\label{final-ImL2}
\end{align}
The evaluation of our imaginary part of the Lagrangian (\ref{final-ImL1}) (\ref{final-ImL2})
is summarized in Fig.\ref{fig:1}.
%%%%%%%%%%%%%%%%%%%%%%%
\FIGURE[t]{ 
\includegraphics[width=6cm]{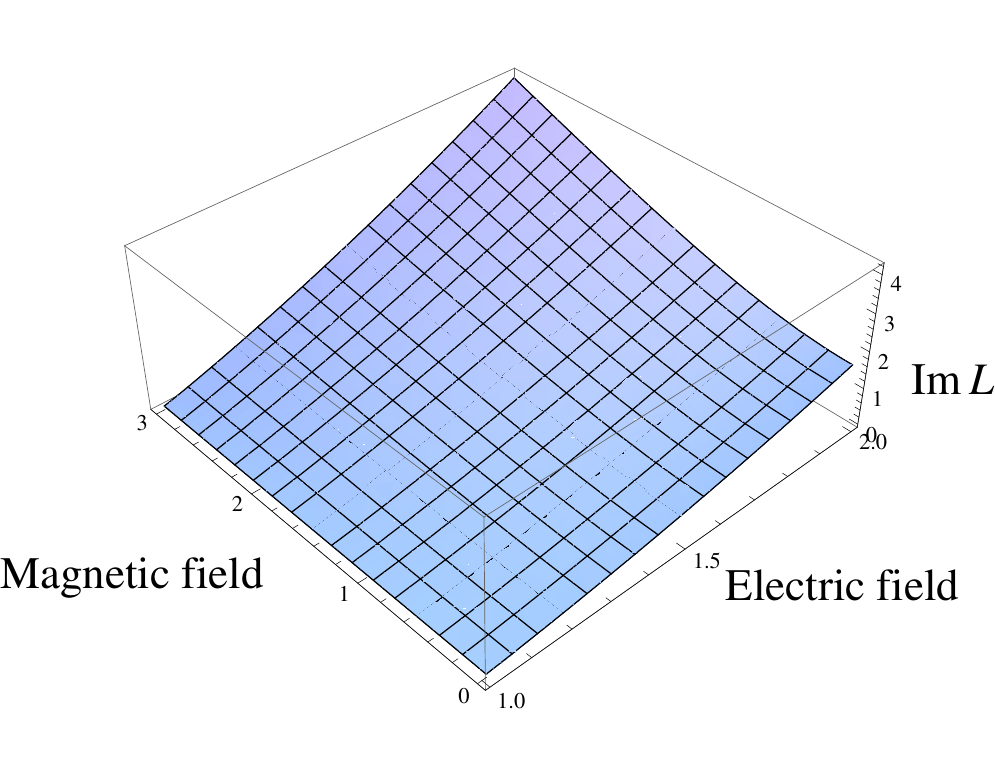}
\includegraphics[width=6cm]{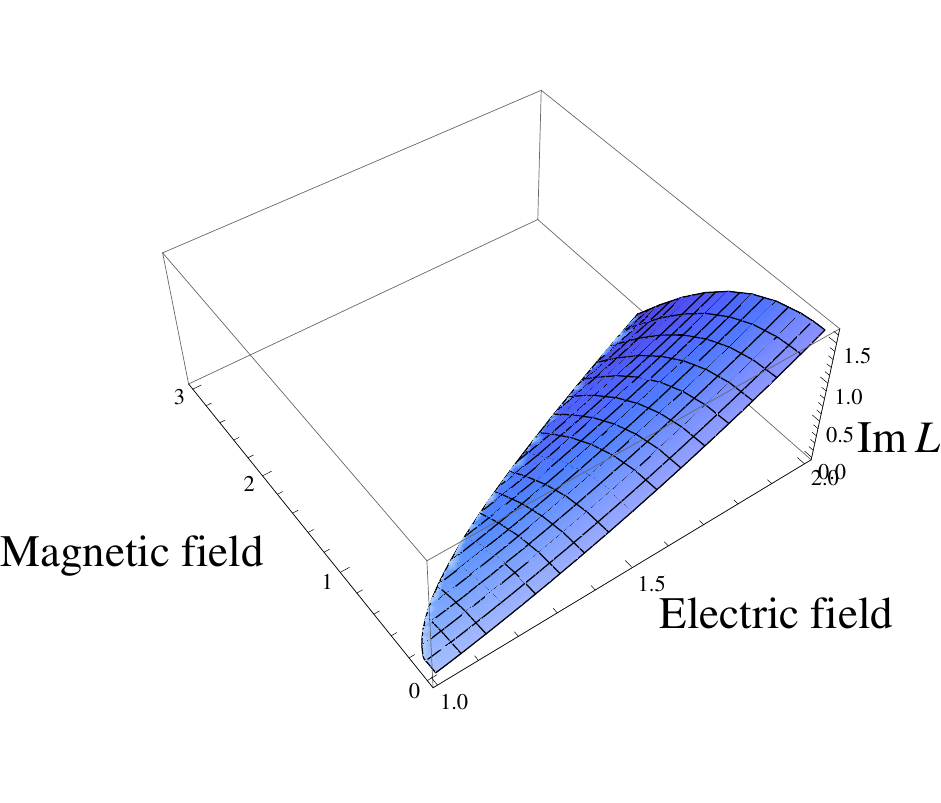}
\caption{The plot of the imaginary part of the Lagrangian. Left: The case with a 
magnetic field parallel to the electric field.
Right: The case with a magnetic field perpendicular to the electric field.
}
\label{fig:1}
}
%%%%%%%%%%%%%%%%%%%%%%%

If we look at only the critical value of the electric field as a function of the magnetic field,
it shows a magnetic catalysis --- the critical electric field only increases once one turns on the magnetic field. The imaginary part of the Lagrangian for the perpendicular magnetic field also follows the magnetic catalysis. However, the imaginary part of the Lagrangian increases
for the parallel magnetic field, which can be interpreted as an inverse magnetic catalysis.
In sum, the behavior of the instability of the system depends on the direction of the magnetic
field relative to the electric field.

%%%%%%

In the next section, we evaluate the imaginary part of the D8-brane action in the deformed Sakai-Sugimoto background.

%%%%%%%%%%%%%%%%%%%%%%%%%%%%%%%%%%%%%%%%%
\section{Pair creation of quark antiquark in deformed D4-D8 brane system}
In this section, in the deformed Sakai-Sugimoto model \cite{Aharony:2006da}, we derive the creation rate of the quark antiquark pair from the imaginary part of the D-brane action with a constant electromagnetic field. We follow a procedure described in the previous section.
\subsection{Euler-Heisenberg Lagrangian of deformed Sakai-Sugimoto model}
In the Sakai-Sugimoto model, the D8-brane and the anti-D8-brane are inserted at the antipodal points of the compactified $S^{1}$, $x^{4}=0$ and $x^{4}=\pi R$.
However, generically $x^{4}$ coordinate for the inserted D-branes can depend on the coordinate $u$, and becomes a function of $u$.
Accordingly, the region of $u$ in which the D8-brane hangs down changes from $[u_{KK}, \infty)$ to $[u_{0}, \infty)$. The D4-brane background is given by (\ref{D4BRANE}).
The coordinate $x^{4}$ of the anti-D8-brane is a function of $u$ and moves in a sub-region of $0<x^{4}(u)<\pi R\hspace{1mm}(u_{KK}<u<\infty)$. When $x^{4}=\pi R\hspace{1mm}(u=u_{KK})$, the model corresponds to the Sakai-Sugimoto model in the previous section. For generic $x^{4}(u)$, the induced metric on the D8-brane is given by
\begin{align}
ds_{D8}^{2} = \left(\frac{u}{R_{D4}}\right)^{3/2}(-dt^2 + \delta_{ij}dx^{i}dx^{j}) + \left(\frac{u}{R_{D4}}\right)^{3/2}\frac{du^{2}}{h(u)}
+ \left(\frac{R_{D4}}{u}\right)^{3/2}u^{2}d\Omega_{4}^{2},
\end{align}
where the region of $u$ is $u_{0}\leq u<\infty\hspace{1mm}(u_{KK}<u_{0}<\infty)$ and the function of $h(u)$ is defined by
\begin{align}
h&(u)\equiv \left[f(u)\left(\frac{dx^{4}(u)}{du}\right)^{2} + \left(\frac{R_{D4}}{u}\right)^{3}\frac{1}{f(u)}\right]^{-1}.
\end{align}

Let us consider the D8-brane action including a constant electromagnetic field in the deformed Sakai-Sugimoto model. Substituting the induced metric on the D8-brane to (\ref{D8DBI}), we obtain the following,
\begin{align}
\mathcal{L} = -\frac{8\pi^{2}}{3}T_{8}\int^{\infty}_{u_{0}}du\frac{u^{4}}{\sqrt{h(u)}}\left(\frac{u}{R_{D4}}\right)^{3/4}\hspace{1mm}e^{-\phi}\sqrt{\xi},
\end{align}
where the function of $\xi$ is defined by
\begin{align}
\xi&\equiv 1 - \frac{(2\pi\alpha')^{2}R_{D4}^{3}}{u^{3}}\left[F_{01}^{2} - F_{12}^{2} - F_{23}^{2} - F_{13}^{2} + h(z)(F_{0u}^{2} - F_{1u}^{2})\right] \notag \\
   & -  \frac{(2\pi\alpha')^{4}R_{D4}^{6}}{u^{6}}\left[F_{01}^{2}F_{23}^{2} + h(u)\{F_{0u}^{2}(F_{12}^{2} + F_{23}^{2} + F_{13}^{2}) - F_{1u}^{2}F_{23}^{2}\} \right]. 
\end{align}
After a massage of the equations, we obtain
\begin{align}
\mathcal{L} = -\frac{8\pi^{2}T_{8}}{3g_{s}}\int^{\infty}_{u_{0}}du\frac{u^{4}}{\sqrt{h(u)}}\sqrt{1 - \frac{(2\pi\alpha')^{2}R_{D4}^{3}}{u^{3}}\left[E^{2}_{1} - \vec{B}^{2}\right] - \frac{(2\pi\alpha')^{4}R_{D4}^{6}}{u^{6}}E_{1}^{2}B_{1}^{2}}, \label{ELDSAKAI}
\end{align}
where the electromagnetic fields are defined by $F_{01}\equiv E_{1},F_{12}\equiv B_{3},F_{23}\equiv B_{1},F_{13}\equiv B_{2}$ and $\vec{B}^{2}\equiv B^{2}_{1}+B^{2}_{2}+B^{2}_{3}$.
\\
\subsection{Imaginary part of the effective action in deformed Sakai-Sugimoto model}

%%%%%%%%%%%%%%%%%%%%%%%
\FIGURE[t]{ 
\includegraphics[width=4.5cm]{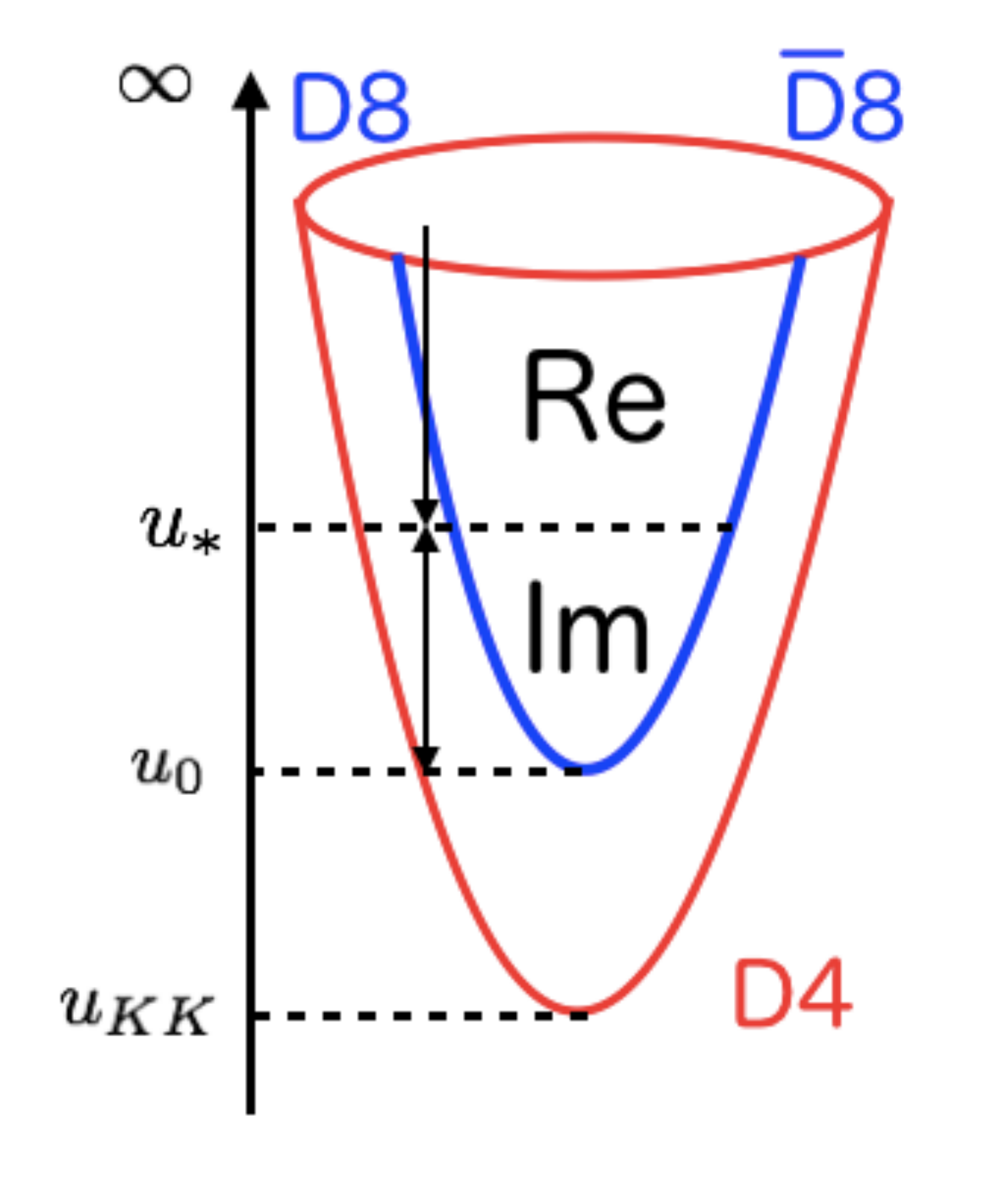}
\caption{
The Euler-Heisenberg Lagrangian has an imaginary part in $u_{0}\le u\le u_{*}$. Since the function of $x^{4}$ depends on $u$ coordinate, the below region of the integral changes from $u_{KK}$ to $u_{0}$.
}
\label{fig:dSS.pdf}
}
%%%%%%%%%%%%%%%%%%%%%%%

In the previous subsection, the D8-brane action in the deformed Sakai-Sugimoto model was obtained as (\ref{ELDSAKAI}). In this subsection, we derive the creation rate of the massless quark antiquark from the imaginary part of the D8-brane action in a constant electromagnetic field.

From (\ref{ELDSAKAI}), we examine the case when the imaginary part of the effective Lagrangian appears. Since the function of $h(u)$ is positive, we should find a region of $u$ such that the square root in the numerator of the integrand has an imaginary part. 
Although the coordinate of $x^{4}$ depends on $u$ in the deformed Sakai-Sugimoto model, the dependence on $u$ in the function of $x^{4}$ has no relation with the imaginary part of the effective Lagrangian. So, we may follow the same logic as given in the previous section.
The condition that this effective Lagrangian has an imaginary part is the same as (\ref{CONDITION}). The integration region of $u$ which gives an imaginary part is

\begin{align}
u_{0}\leq u < \left[\frac{(2\pi\alpha')^{2}R^{3}}{2}\left\{ E^{2}_{1} - \vec{B}^{2} + \sqrt{(E^{2}_{1} - \vec{B}^{2})^{2} + 4E_{1}^{2}B_{1}^{2}}\right\} \right]^{1/3}. \label{CEDSAKAI}
\end{align}
The imaginary part of the effective Lagrangian is evaluated as 
\begin{align}
\mathrm{Im}\mathcal{L} = \frac{8\pi^{2}T_{8}}{3g_{s}}\int^{u_{*}}_{u_{0}}du\frac{u^{4}}{\sqrt{h(u)}}\sqrt{\frac{(2\pi\alpha')^{4}R^{6}}{u^{6}}E_{1}^{2}B_{1}^{2} + \frac{(2\pi\alpha')^{2}R^{3}}{u^{3}}\left[E^{2}_{1} - \vec{B}^{2}\right] - 1},
\end{align}
where $u_{*}$ is defined by (\ref{U}). The integral region of $u$ is shown in Fig. \ref{fig:dSS.pdf}

Next, we evaluate the critical electric field. The critical electric field is derived from the condition that the imaginary part of the effective Lagrangian starts to grow. From (\ref{CEDSAKAI}), we obtain
\begin{align}
u_{0} \leq \left[\frac{(2\pi\alpha')^{2}R^{3}}{2}\left\{ E^{2}_{1} - \vec{B}^{2} + \sqrt{(E^{2}_{1} - \vec{B}^{2})^{2} + 4E_{1}^{2}B_{1}^{2}}\right\} \right]^{1/3}.
\end{align}
The critical electric field $E_{\mathrm{cr}}$ is obtained by the following,
\begin{align}
E_{\mathrm{cr}} = \left[\frac{u^{3}_{0}}{(2\pi\alpha')^{2}R^{3}}\cdot\frac{\left\{\frac{u^{3}_{0}}{(2\pi\alpha')^{2}R^{3}} + \vec{B}^{2}\right\}}{\left\{\frac{u^{3}_{0}}{(2\pi\alpha')^{2}R^{3}} + B_{1}^{2}\right\} } \right]^{1/2}.
\end{align}
This critical electric field is of the same form as that for the critical electric field in the Sakai-Sugimoto model, if we change from $u_{KK}$ to $u_{0}$ on (\ref{CRESAKAI}). 
Indeed, this expression coincides with the generic formula (\ref{EcrBst}), though
the parameter appearing here is different from the QCD string tension of the D4-brane
geometry $(2/27)\lambda M_{\rm KK}^2$. This is because the D8-brane does not reach the bottom of the confining geometry and does not satisfy the assumption to derive the generic formula (\ref{EcrBst}) with the QCD string tension.
When $B_{2},B_{3}=0$, the critical electric field is $E_{\mathrm{cr}}=\left[u_{0}^{3}/(2\pi\alpha')^{2}R^{3}\right]^{1/2}$, which is the independent of $B_{1}$ as in the case of the Sakai-Sugimoto model.

%%%%%%%%%%%%%%%%%%%%%%%%%%%%%%%%%%%%%%%%%

\section{Summary}
In this paper, we studied the vacuum instability induced by a constant electromagnetic field by evaluating the Euler-Heisenberg Lagrangian of the large $N_{c}$ non-supersymmetric QCD with the (deformed) Sakai-Sugimoto model in the gravity side. Since the Sakai-Sugimoto model has a confining scale, we obtained qualitatively different results from that of $\mathcal{N}=2$ SQCD \cite{Hashimoto:2014dza}.

By evaluating the imaginary part of the Euler-Heisenberg Lagrangian in the large $N_{c}$ QCD, we found that the creation rate of the massless quark antiquark is finite as oppose to the results in the $\mathcal{N}=2$ SQCD.
We found that the imaginary part of the Euler-Heisenberg Lagrangian increases when the magnetic field parallel to the electric field increases, on the other hands, it deceases when the magnetic field perpendicular to the electric field.
We also obtained the critical electric field by the condition such that the effective Lagrangian has an imaginary part. It was shown to have the universal form. 

There are several issues concerning the instability of the holographic QCD set-up we used in this paper. We found an instability caused on the flavor D-brane by the electric field. 
How the instability results in a dynamical decay process is beyond our scope of this paper.
In fact, since the background geometry is a confining geometry, if we keep the geometry
during the decay, it is impossible to have an electric current --- there is no place for the
flavor D-brane to end in the geometry, as opposed to the situation with a black hole horizon in the bulk (for example in the case of supersymmetric QCD). To make an electric current flow, one may need a baryon vertex, but generically it is too heavy to create. So, we are not sure where the dynamical instability leads us to. Creation of such a baryon vertex in a time-dependent holographic QCD is an interesting question. We leave it to a future work.

%%%%%%%%%%%%%%%%%%%%%%%%%%%%%%%%%%%%%%%%

\section*{Acknowledgments}

K.H. would like to thank K.-Y.~Kim, Y.~Sato and K.~Yoshida. 
A.S. would like to thank T.~Enomoto, Y.~Hidaka and S.~Yamaguchi.
This research was
partially supported by the RIKEN iTHES project.

%%%%%%%%%%%%%%%%%%%%%%%%%%%%%%%%
%%%%%%%%%%%%   Reference
%%%%%%%%%%%%%%%%%%%%%%%%%%%%%%%%

\include{begin}

\include{end}

\end{document}